\pdfoutput=1

\documentclass[a4paper, 11pt]{article}

\usepackage[T1]{fontenc} 
\usepackage[utf8]{inputenc} 
\usepackage{amsmath} 
\usepackage{calc} 
\usepackage[hmargin=3cm, vmargin=2.5cm]{geometry} 
\usepackage[hidelinks]{hyperref} 

\usepackage{mathpazo} 
\usepackage{helvet} 
\usepackage{bm} 

\usepackage[font={sf, small}]{caption}
\usepackage[sf, bf, medium]{titlesec}

\usepackage[square,comma,sort&compress,numbers]{natbib}
\usepackage{graphicx}		
\DeclareGraphicsExtensions{.png}	
\usepackage{float}
\usepackage{subcaption}

\usepackage{fancyhdr}
\begin{document}

\null
\begin{center}
\sffamily \LARGE \bfseries

Characterization of ultrasound transmit-receive measurement systems in air\\
\normalsize {Comparison with prior work on piezoelectric elements in radial mode vibration}

\large \mdseries \vspace{\baselineskip}

Renate Grindheim\textsuperscript{1}, 
Per Lunde\textsuperscript{1,2}, 
Magne Vestrheim\textsuperscript{1} \\ 

\small \vspace{\baselineskip}

\textsuperscript{1} University of Bergen, Department of Physics and Technology, P.O. Box 7803, N-5020 Bergen, Norway \\ 
\textsuperscript{2} NORCE, Norwegian Research Centre AS, P.O. Box 6031, N-5898 Bergen, Norway \\ 
\newcommand{\email}{renate@grindheim.net} 
Contact email: \href{mailto:\email}{\email}

\end{center}
\vspace{\baselineskip}

\noindent\hspace*{\fill}\begin{minipage}{\textwidth-2cm}
{\large \sffamily \centering \textbf{Abstract} \\} \small

A system model is used, describing the voltage-to-voltage transfer function for an ultrasonic transmit-receive transducer pair based on simulations using a finite element model for piezoelectric transducers (FEMP). The piezoelectric ceramic elements used in this work are cylindrical Pz27 disks vibrating radially in air at room temperature and 1 atm, over a frequency range up to 300 kHz. Comparisons are made between measurements and simulations. Effects of uncertainties in alignment are examined, together with comparisons with prior work. 
\end{minipage}\hspace*{\fill}
\vspace{\baselineskip}

\section{Introduction}
Accurate modelling of ultrasonic measurement systems for characterization of gas is of interest in industrial and scientific applications. Examples include fiscal flow measurement for custody transfer of natural gas \citep{ISO, FroysaLunde2003}, energy and quality measurement of gas \citep{FroysaLunde2003, FroysaLunde2006, Farz}, and sound velocity and absorption measurements \citep{Norli, Lunde2007}. In the present work a “system model” of  a transmit-receive measurement system refers to “a mathematical / numerical model aiming to describe the chain of electro-acoustic signal propagation through the system, from the electrical signal generator to the electrical recording equipment (e.g., an oscilloscope), via the piezoelectric transmit and receive transducers, the propagation medium, and possible transmit and receive cables/electronics” \citep{rune}.  
	\\ \indent Various approaches have been used to describe transmit-receive ultrasonic measurement systems for fluids. This includes methods based on (i) the Mason model [or equivalent one-dimensional (1D) approaches] for thickness-mode transmitting and receiving transducers, combined e.g. with uniform piston \citep{FroysaLunde2003, Stepanishen81, Lygre87, Collie89, Vervik95, Willatzen99, Vervik2000, Willatzen2001} or plane wave \citep{Papadakis77, Low80, Hayward84, Hayward842, Yamaguchi86, Wilcox98} type of radiation models; (ii) 1D “electroacoustic measurement (EAM) model” types of description \citep{Dang02, Dang022, Schmerr07, Schmerr11}; and (iii) purely electrical transmission line system modelling \citep{Deventer00}. More rigorous and accurate descriptions based on finite element modelling (FEM) have proven highly useful for ultrasonic system modelling, accounting for e.g. thickness-mode \citep{Lunde03, Bezdek07, Bezdek08, Bezdek082, Ge14} and radial mode \citep{Norli, Hauge, Mosland, Storheim, Sovik, Andersen, Hagen, rune} transducers in such measurement setups. 
\\ \indent The work presented here builds on prior work described in Refs. \cite{Norli, Mosland, Sovik, Hauge, Andersen, Hagen, rune}, using finite element modelling of an ultrasonic measurement system for air employing piezoelectric elements vibrating in their lower radial modes. Measurements and simulations of the frequency response of a voltage-to-voltage transmit-receive transfer function are compared, including comparison with some recent work in this area \citep{Andersen, Hagen}. A second objective is to investigate and quantify measurement sensitivities to uncertainties in positioning and alignment of the transmitting and receiving piezoelectric elements in this setup.

\section{Theory}

The experimental measurement setup used in the present work is described in Section 3.1. A system model based on this setup is used. This was first introduced by \citep{Hauge, Mosland}, cf. \citep{rune}, and further developed by \citep{Sovik, Andersen, Hagen}. 

\subsection{System model}
\begin{figure}[H]
\centering\noindent\makebox[\textwidth]
{\includegraphics[width=0.7\paperwidth]{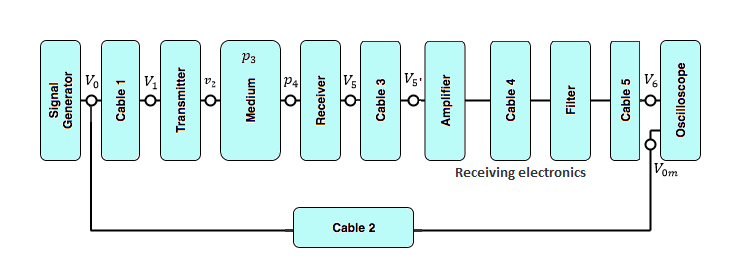}}
\caption{Block diagram representation of the system model used in the present work.}\label{block}
\end{figure}

The system model illustrated in Figure \ref{block} shows the different components included in the measurement setup as linear blocks where transfer functions relate the signal going from one block to another in the frequency domain. The system model consists of two cylindrical, 10 mm x 2 mm, Pz27 piezoelectric ceramic disks in radial mode vibration in air, where the transmitting disk, $T_x$, is coupled to the transmitting electronics (signal generator, oscilloscope and cables), and the receiving disk, $R_x$, is coupled to the receiving electronics (amplifier, filter, cables and oscilloscope). The two disks are mounted parallel (with respect to the $xy$-plane, cf. Figure \ref{fig:1a}a) at a separation distance $d$ along the $z$-axis. 

The voltage signals $V_6(t)$ and $V_{0m}(t)$ measured at the oscilloscope are Fourier transformed as described in \citep{Hagen}, giving the respective spectral components $V_6(f)$ and $V_{om}(f)$, where $t$ is the time and $f$ is the frequency.
For these spectral components a monochromatic time dependency $e^{i\omega t}$ is assumed and suppressed. At angular frequency $\omega = 2\pi f$ $V_0(f)$ is the output voltage spectral component generated by the signal generator and $V_{0m}(f)$ is the output voltage spectral component measured in Channel 1 on the oscilloscope.
$V_1(f)$ is the input voltage spectral component at the transmitting disk, $T_x$.
$v_2$ is the spectral component of the particle velocity at the center of the front surface of $T_x$.
$p_3$ is defined in \citep{Mosland} and not discussed in the current work. $p_4$ is the on-axis spectral component of the free-field pressure at the front of $R_x$.
$V_5(f)$ is the output voltage spectral component at $R_x$, and $V_{5'}(f)$ is the output voltage spectral component at the amplifier.
$V_6(f)$ is the input voltage spectral component measured and terminated in Channel 2 on the oscilloscope. 

\subsection{Transfer functions}
The voltage spectral components $V_6(f)$ and $V_{om}(f)$ are divided to obtain the voltage-to-voltage transfer function of the complete signal chain as defined in Equation (2.1) in \citep{Sovik} with the receiving electronics treated as in Equation (2.64) in \citep{Hagen}, i.e.

\begin{equation}
\label{H06}
H_{0m6}^{VV}(f) \equiv \frac{V_6(f)}{V_{0m}(f)} = \frac{V_1(f)}{V_{0m}(f)}\cdot \frac{V_{5open}(f)}{V_{1}(f)}\cdot \frac{V_{5'}(f)}{V_{5open}(f)}\cdot \frac{V_{6}(f)}{V_{5'}(f)}.
\end{equation}

The various transfer functions used in Eq. (1) are described in the following.
The open-circuit voltage-to-voltage transfer function describing the sound propagation from $T_x$ to $R_x$ at lossless conditions in the air medium is defined as equation (2.20) in \citep{Hauge},

\begin{equation}
\label{H15opendef}
H_{15open}^{VV}(f) = |H_{15open}^{VV}(f)| e^{i\theta_{15open}} \equiv \frac{V_{5open}}{V_1},
\end{equation}
where $V_{5open}$ is the open-circuit voltage at $R_x$. $|H_{15open}^{VV}(f)| $ is the magnitude and $\theta_{15open}$ the phase of this transfer function. The phase can be decomposed into a slowly varying phase and a plane wave component, cf. Equation (2.100) in  \citep{Sovik} and Equation (2.14) in \citep{Andersen}. The slowly varying phase is then given by

\begin{equation}
\label{thetaH15open}
\theta_{15open}^{slow}=\theta_{15open} +kd,
\end{equation}
where $k=\omega /c$ is the wave number and $c$ is the speed of sound in air. 

The transfer function relating the measured voltage at Channel 1 on the oscilloscope to the voltage at $T_x$, $H_{0m1}^{VV}$, is defined as equation (2.5) in \citep{Andersen} 
	
	\begin{equation}
\label{H0m1}
H_{0m1}^{VV}(f) = \frac{V_1(f)}{V_{0m}(f)}.
\end{equation}

\noindent The transfer functions $H_{5open5}^{VV}$, describing the influence of the electrical load from cable 3 and the receiving electronics on the output voltage from $R_x$, and $H_{5'6}^{VV}$, describing the receiving electronics including cables 4 and 5, are defined as \citep{Hagen},

	\begin{equation}
\label{H5open5}
H_{5open5'}^{VV}(f) =\frac{V_{5'}(f)}{V_{5open}(f)}
\end{equation}

 \noindent and
	
 \begin{equation}
\label{H56}
H_{5'6}^{VV}(f) = \frac{V_6(f)}{V_{5'}(f)},
\end{equation}

\noindent respectively.

\subsection{Simulated transfer function $H_{15open}^{VV}$}
The simulated lossless open-circuit transfer function $H_{15open}^{VV}$ is calculated by assuming spherical reciprocity and using the simulated far-field axial pressure $p_{ff}=p_4(z_{ff})$, where $z_{ff}=1000$ m is used, extrapolating it to the desired distance, $d$. This model assumes identical transducers and does not include losses in the medium or diffraction effects caused by the distance, $d$, being in the near-field of the transducers. The lossless open-circuit transfer function is then given as Equation (2.52) in \citep{Andersen},

\begin{equation}
\label{H15opensim}
H_{15open}^{VV}(f)= \frac{Z_Tp_4^2(z_{ff})2z_{ff}^2}{i\rho d f } e^{ik(2z_{ff}-d)},
\end{equation}

\noindent where the input electrical impedance of $T_x$, $Z_T$, and the far-field pressure, $p_4(z_{ff})$, is simulated using finite element modelling (FEMP) \citep{Kochbach}, and $\rho$ is the density of the medium. 

\subsection{Measured transfer function $H_{15open}^{VV}$}
In order to obtain a lossless open-circuit transfer function, $H_{15open}^{VV}(f)$, from measurements, corrections are made for attenuation in the medium and near-field effects caused by diffraction due to the finite sizes of $T_x$ and $R_x$.The attenuation correction factor, $C_\alpha$, is found by accounting for attenuation due to the classical absorption of sound in air, $\alpha_{cl}$, rotational motion of the air molecules, $\alpha_{rot}$, vibrations of oxygen molecules, $\alpha_{vib,O}$ and vibration of nitrogen molecules, $\alpha_{vib,N}$. These are calculated according to \cite{SI1.26} with the theory for implementation in the system model described in \citep{Mosland}. The diffraction correction $C_{dif}$, is found by using a baffled piston diffraction correction factor as in \citep{Hauge, Storheim, Sovik, Mosland, Andersen, Hagen}, where the diffraction correction is calculated according to Khimunin's model \citep{Khimunin72, Khimunin75}. 

$H_{0m1}^{VV}$ and $H_{5open5'}^{VV}$ are estimated using transmission line models as proposed in \citep{Sovik}. $H_{5'6}^{VV}$ is found by measuring the transfer function of the receiving electronics including cables 4 and 5, by coupling the signal generator to an attenuator and then directly into the receiving electronics, bypassing the transmitting and receiving transducer completely, as in \citep{Hagen}. 

From Equation \ref{H06}-\ref{H56} and the corrections mentioned above, the lossless transfer function $H_{15open}^{VV}(f)$ is calculated as 

\begin{equation}
\label{H15openf}
H_{15open}^{VV}(f) = \frac{H_{0m6}^{VV}(f)}{H_{0m1}^{VV}(f)H_{5open5'}^{VV}(f)H_{5'6}^{VV}(f)} C_\alpha C_{dif}.
\end{equation}

\section{Methods}
The experimental setup has been largely kept as used by \cite{Sovik, Andersen, Hagen} with only minor changes such as the lengths of cables 4 and 5, and the different instruments having been moved due to a renovation at the lab in 2017. 

\subsection{Experimental Setup}
The two piezoelectric disks are mounted in air on positioning stages (Physik Instrumente GmbH$\&$Co, Germany). $T_x$ is connected to a PI linear positioning stage (PI M-531.DG \citep{MP531}), allowing movement in the z-direction, and to a PI rotational stage (PI M-037.PD \citep{M037}) for rotation about the $x$-axis, cf. Figure \ref{fig:1a}. $R_x$ is connected to a PI linear position stage (PI M-535.22 \citep{MP33E}) moving in $y$-direction, and can be rotated about the $x$-axis by loosening a screw at the top of the rod it is connected to, allowing it to be turned. Two lasers, mounted on a rod, can be placed between the two disks in order to measure (i) the alignment and (ii) the distance, $d$ (cf. Figure \ref{fig:1a}b). In this work the distance $d$ = 0.50 m, is used. 
\\ \indent An Agilent 33220A function generator is used to generate the input signal at desired voltage over the frequency range 30 - 300 kHz. The signal $V_{0m}(t)$ is then measured using a Tektronix DPO3012 digital oscilloscope. $R_x$ is connected to a Br{ü}el$\&$Kjær 2636 amplifier and a Krohn-Hite 3940A digital filter before the signal terminates in the Tektronix DPO3012 digital oscilloscope and $V_6(t)$ is measured. The cables used to connect the instruments (cables 2, 4 and 5) are coaxial cables of type RG-58, with lengths $l_2 = 0.280$ m, $l_4 = 0.475 $ m and $l_5 = 1.470 $ m. The cables used to connect the oscilloscope to $T_x$ and $R_x$ to the amplifier (cables 1 and 3, respectively) are coaxial cables of type RG-178 B/U, with length $l_1 = 2.970$ m and $l_3 = 2.975 $ m, respectively. For the present work the piezoelectric disk elements used are denoted elements 7 and 13, as transmitter and receiver, respectively. This is the same as used in \citep{Hagen}, whereas \citep{Andersen} used elements 4 and 13, cf. the comparisons made in Section 4.2. 

\begin{figure}[H]
\centering\noindent\makebox[\textwidth]
{\includegraphics[width=0.4\paperwidth]{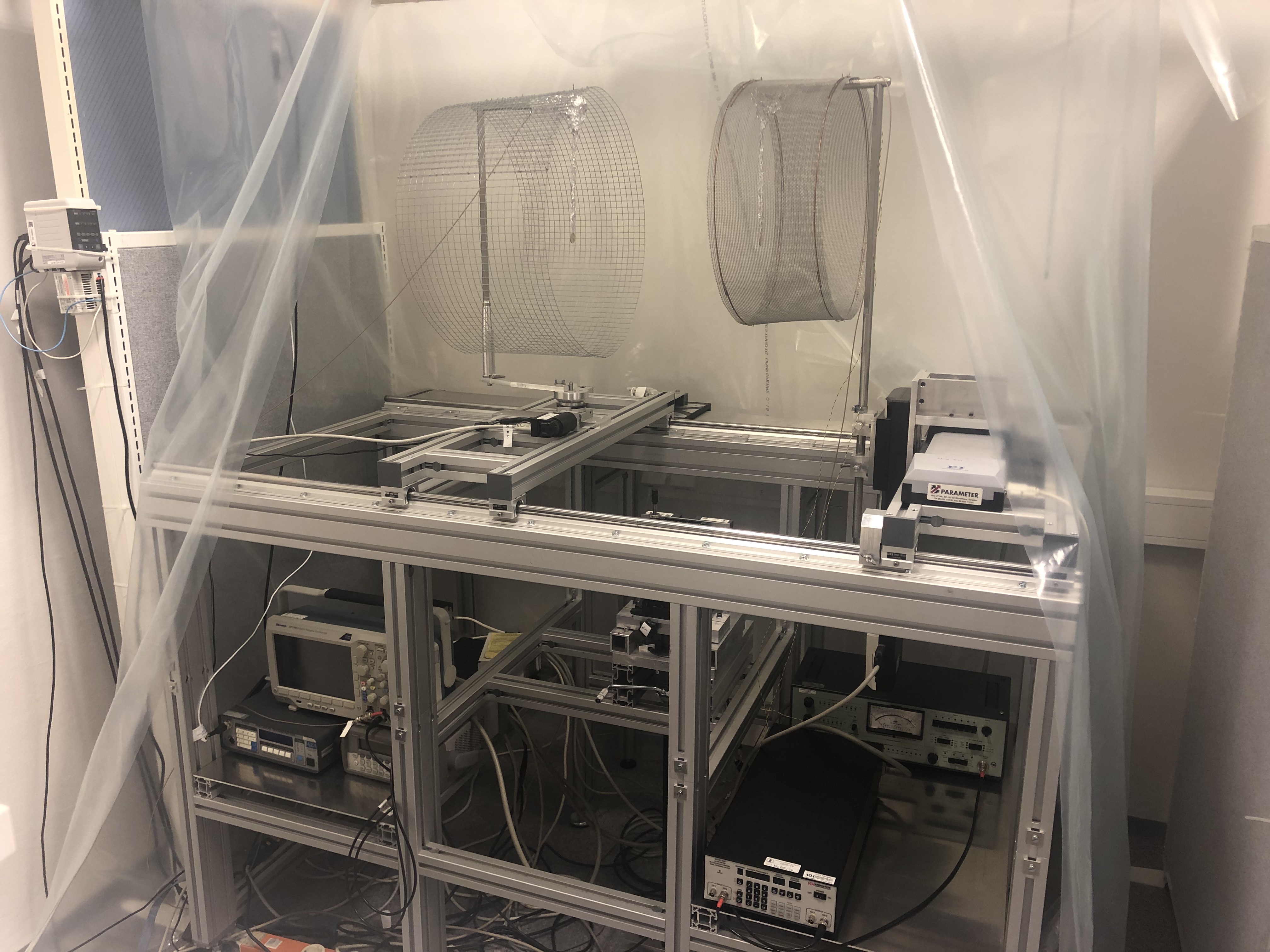}}
\caption{Experimental setup in the acoustics laboratory.}\label{lab}
\end{figure} 

\subsection{Alignment of the transmitter, $T_x$, and receiver, $R_x$.}
\begin{figure}
\begin{subfigure}{0.31\textwidth}
\includegraphics[width=0.23\paperwidth]{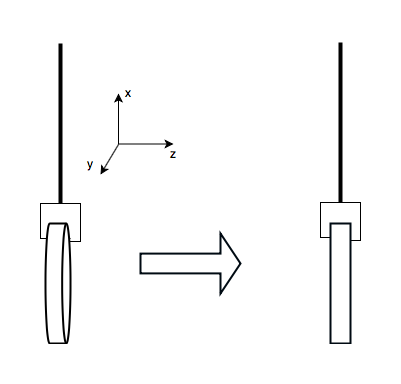}
\end{subfigure}
\hspace{30mm} 
\begin{subfigure}{0.31\textwidth}
\includegraphics[width=0.3\paperwidth]{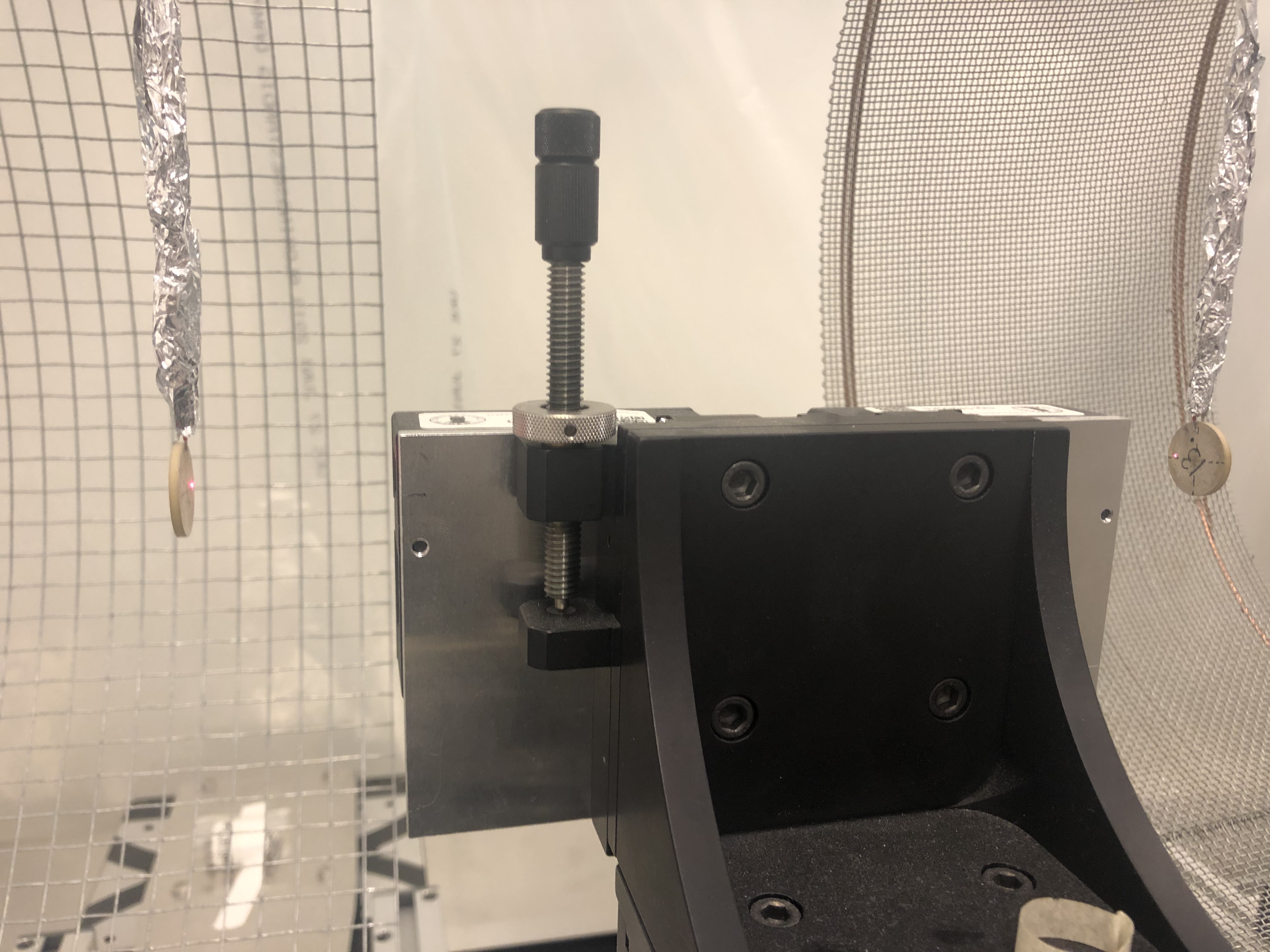}
\end{subfigure}
\caption{Sketch of a) the mounting of the piezoelectric ceramic disks and b) aligning of the disks using laser.} \label{fig:1a}
\end{figure}

The disks are soldered onto partially twisted pairs cables, shielded with aluminium foil, mounted in a Faraday cage, cf. Figures \ref{lab} and \ref{fig:1a}b. The aim is to have the disks centred on the $z$-axis and place them parallel to the $xy$-plane as shown in Figure \ref{fig:1a}. This is done using two high precision Keyence lasers \citep{Andersen}, measuring differences in the + and - $z$-direction to measure the position of the disks, and the high precision translation stages from Physik Instrumente, cf. Section 3.1, to position the disks.
\\\indent Firstly the disks are roughly placed on the $z$-axis by adjusting $R_x$ in $x$- and $y$-direction so that the two laser beams point to the x marking the center on each of the disks. This mark has a finite size, so the laser is moved to the top, bottom (vertically) and side edges (horizontally) of the disks to see if these are aligned. Adjustments are made by moving $R_x$ in $x$- and $y$-direction if necessary. 
\\\indent When the disks are both centred on the $z$-axis and aligned vertically and horizontally, the rotational alignment is investigated. For rotation about the $y$-axis this is done by moving the laser to the  top and the bottom of $T_x$, noting the distance from the laser to the disk in both positions and subtracting them to find the difference. This is referred to as uncertainty in the alignment caused by rotation about the $y$-axis. If alignment is not satisfactory the disk is moved by a light touch at the bottom of the disk, continuously monitoring the position on the laser display until the desired deviation about the $y$-axis is acquired.  This procedure is then repeated for $R_x$.
\\\indent For rotation about the $x$-axis the laser is placed on one of the furthest horizontal sides of the disk. The uncertainty in alignment is found by measuring at the two horizontal sides of the disk and subtracting the two distances to find the difference. For $T_x$ corrections are now made by using the rotation stage mentioned in Section 3.1. For $R_x$ a light touch or a screw at the top of the mounting rod allows for the rod and the disk to be rotated and is used to correct its positioning. 
\\\indent The process is repeated several times, to check that corrections in one direction did not affect the deviation or uncertainty in alignment in another direction, until satisfactory alignment is achieved.
\\\indent Lastly the positioning stage in $z$-direction is used to position the front of the transmitting disk at the separation distance $z = d$ and the front face of the receiver is placed at $z = 0$. The combined standard uncertainty for the separation distance, $d$, as claimed in Section 7.1.1 in \citep{Andersen}, is 40 $\mu$m at a coverage factor $k$ = 1.  

The small and bright red/white area on each piezoelectric disk in Figure \ref{fig:1a}b shows the "footprint" of the laser beam on the disk, for a given position of the laser, at which a distance measurement is made. As described above, the laser beam is then traversing over the disk surface in a "diametrical cross formation" (in the $x$- and $y$-directions), to get a distribution of distance measurements over this cross formation.


\section{Results}
\subsection{The influence of accuracy in alignment }

\begin{figure}[H]
\centering\noindent\makebox[\textwidth]
{\includegraphics[width=0.65\paperwidth]{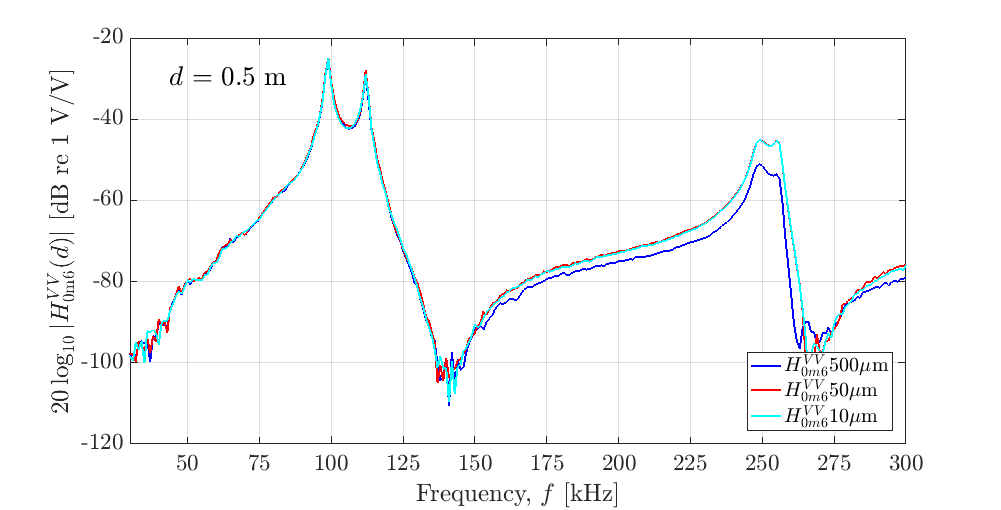}}
\caption{Measurements of the magnitude of the transfer function $H_{0m6}^{VV}$ for different accuracies in rotational alignment of the disks.}\label{aligning}
\end{figure}

In \citep{Andersen} the tolerated uncertainty in alignment of the disks in the $xy$-plane was given as 10 $\mu$m or less, whereas \citep{Hagen} allowed for uncertainties up to 100 $\mu$m. Therefore the effect of different alignment uncertainty is investigated by aligning the disks to different degrees of accuracy and observing how this affects the transfer function, $H_{0m6}^{VV}$. The accuracies chosen were 500 $\mu$m, 50 $\mu$m and 10 $\mu$m. The results are shown in Figure \ref{aligning}. 

For the first two peaks, associated with the fundamental radial mode $R_1$, there is little to no change in the magnitude of the transfer function $H_{0m6}^{VV}$, for all three accuracies. At the second pair of peaks, associated with the second radial mode $R_2$, there is no noticeable change in the magnitude for alignment to the degree of 50 $\mu$m and 10 $\mu$m. However, the measurement with 500 $\mu$m uncertainty in the alignment shows a decrease in magnitude of almost 5 dB compared to the measurements with 50 $\mu$m and 10 $\mu$m misalignment.

The effect of misalignment in the slowly varying phase will be investigated further in \citep{Grindheim}.

\subsection{Comparison with prior work}
\begin{figure}[H]
\centering\noindent\makebox[\textwidth]
{\includegraphics[width=0.65\paperwidth]{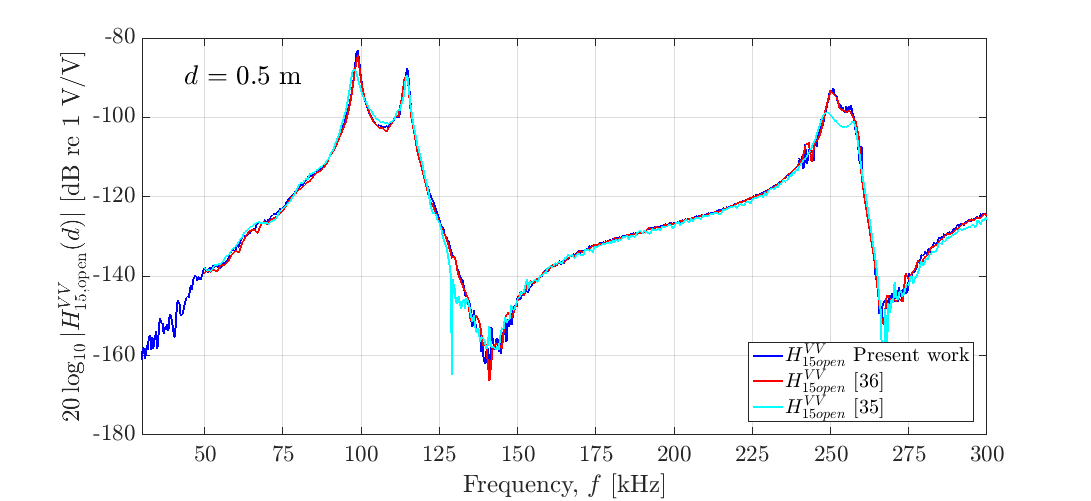}}
\caption{Measurements of the magnitude of the transfer function $H_{15open}^{VV}$ as compared with prior work \citep{Andersen, Hagen}.}\label{sammen}
\end{figure}

Measurements of the open-circuit transfer function $H_{15open}^{VV}$ as compared to prior work \citep{Andersen, Hagen} is shown in Figure \ref{sammen}. Comparisons of the slowly varying phase is also of interest  and will be further discussed in \citep{Grindheim}.

It should be noted that the receiving disk used by Hagen \citep{Hagen} and in the present work are not the same as used by Andersen \citep{Andersen}, and the calculation of the receiving electronics transfer function, $H_{5'6}^{VV}$, differ slightly as the filter is coupled in series after the amplifier in the present work and \citep{Hagen}, whereas \citep{Andersen} coupled the filter between the input and output sections of the amplifier. The signal generator voltage settings used for measurements vary for the different frequencies, cf. \citep{rune}. At $R_1$ the voltage settings used are 0.1 V for the present work and 1 V for \citep{Andersen} and \citep{Hagen}. For $R_2$ the voltage settings used are 0.1 V in the present work and \citep{Hagen} and 1 V in \citep{Andersen}. For all other frequencies the voltage setting is 10 V in the current work, \citep{Andersen} and \citep{Hagen}. The frequency steps used in the present work, \citep{Andersen} and \citep{Hagen}, are 300 Hz, 500 Hz and 1 kHz, respectively. 

In the frequency range 30 - 80 kHz the measurements in the present work reveal a clear undulation pattern, which is also present in the measurements presented in \citep{rune, Hauge, Mosland} (but not as clear in \citep{Andersen, Hagen}). In \citep{Storheim} these are shown to be due to side radiation from $T_x$. 

For the magnitude of $H_{15open}^{VV}$ at the first peak associated with $R_1$, the deviations of the present work are about 4.5 and 1 dB relative to \citep{Andersen} and \citep{Hagen}, respectively. For the second peak associated with $R_1$ the deviations of the present work are 1.5 and 1 dB, relative to \citep{Andersen} and \citep{Hagen}. 

Similarly, for $R_2$, the corresponding deviations of the present work at the first peak are 6 and 1 dB, relative to \citep{Andersen} and \citep{Hagen}. At the second peak associated with $R_2$, the deviations of the present work relative to \citep{Andersen} and \citep{Hagen} are 4 and 1 dB, respectively. 

The oscillations in the magnitude of the current work and \citep{Hagen} about 240 kHz is likely due to the switch from 10 V input voltage to 0.1 V input voltage for the $R_2$ frequency range. This will be investigated further in \citep{Grindheim}.

With respect of the frequencies for the two peaks associated with $R_1$ and the first peak associated with $R_2$, the corresponding deviations of the present work relative to \citep{Andersen} and \citep{Hagen} are about 1 kHz, typically. At the second peak associated with $R_2$, the deviations of the present work relative to \citep{Andersen} and \citep{Hagen} are 3.5 and 2 kHz, respectively. 

\begin{figure}[H]
\centering\noindent\makebox[\textwidth]
{\includegraphics[width=0.65\paperwidth]{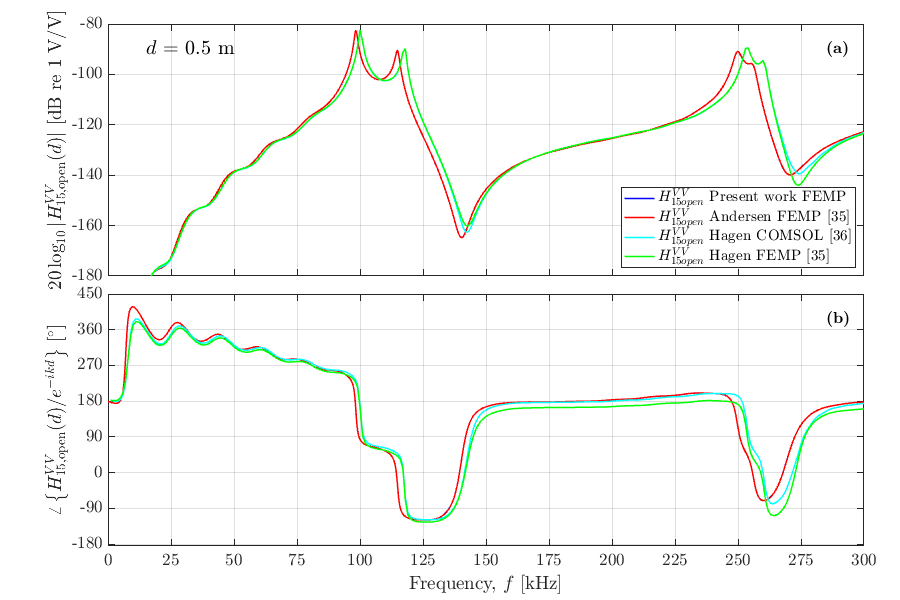}}
\caption{Simulated transfer function $H_{15open}^{VV}$  for a) the magnitude and b) the slowly varying phase, $\theta_{15open}^{slow}$, as compared with prior work.}\label{sammensim}
\end{figure}

Figure \ref{sammensim} shows simulations of the magnitude and the slowly varying phase of the transfer function $H_{15open}^{VV}$, as compared to prior work \citep{Andersen, Hagen}, calculated using two different finite element software tools: FEMP \citep{Kochbach} and COMSOL \citep{comsol}. 

The same material parameters, piezoelectric disk dimensions, software (FEMP) and finite element mesh, are used for simulations in the current work and by Andersen \citep{Andersen}. Therefore these curves are identical, cf. Figure \ref{sammensim}. The simulations done in COMSOL and FEMP by Hagen \citep{Hagen} differ from the current work and Andersen \citep{Andersen}. There is a lower magnitude found in the present work and \citep{Andersen} compared to \citep{Hagen} for the first and second peak associated with $R_1$, of about 0.3 dB  and 0.7 dB, respectively. For the two peaks associated with $R_2$ the deviations are about 0.6 dB  and 1 dB, respectively. There is a lower frequency for the two peaks associated with $R_1$ for the present work and \citep{Andersen} as compared to \citep{Hagen}, and the differences are about 2 and 3 kHz, respectively. For the two peaks associated with $R_2$, the differences are about 3 and 5 kHz, respectively.


\section{Conclusions and further work}
The sensitivity of the transfer function $H_{15open}^{VV}(f)$ with respect to rotary alignment about the $y$- and $x$-axes is investigated. It is found that a rotary misalignment less than 50 $\mu$m does not significantly influence on $|H_{0m6}^{VV}(f)|$. However a rotary misalignment $\approx$ 500 $\mu$m reduces $|H_{0m6}^{VV}(f)|$ significantly, especially for $R_2$, at which there is a decrease of approximately 5 dB. 
\\\indent Comparisons with prior experimental measurements show some deviations from \citep{Andersen} and \citep{Hagen}. The largest deviations are found in the first peak associated with $R_1$ and at the two peaks associated with $R_2$. For $R_2$ an input voltage of 1 V was applied by \citep{Andersen}, whereas \citep{Hagen} used 0.1 V, as in the current work. The low voltage of 0.1 V at $R_2$ and $H_{5'6}^{VV}(f)$ as defined in \citep{Hagen} may explain some of the deviations and will be further investigated  in \citep{Grindheim}.
\\\indent Comparisons with prior simulations \citep{Andersen, Hagen} show deviations from \citep{Hagen}. Both higher frequency for the four peaks of $R_1$ and $R_2$ (2, 3, 3 and 5 kHz, respectively) and larger magnitude at $R_2$ (0.6 and 1 dB, for the two peaks) are observed in the simulations of \citep{Hagen} as compared the current work and \citep{Andersen}. The simulations by \citep{Hagen} using FEMP and COMSOL show some deviation, especially above $R_2$. The material data, the dimensions of the simulated piezoelectric disk elements chosen and the finite element mesh chosen may contribute to the observed deviations. This will be investigated further in \citep{Grindheim}. 

\section{Acknowledgements}
The present work is part of an ongoing master work on the sound field and measurement system modelling using piezoelectric elements vibrating in air, due by June 2019 \citep{Grindheim}. The following persons are acknowledged for guidance and assistance: Espen Storheim, Andreas Hagen, Magnus Rentch Ersdal and Kenneth K. Andersen. 

\bibliography{SSPA19-article}
\end{document}